\documentclass[showpacs,superscriptaddress,pre,twocolumn]{revtex4-1}%
\usepackage{graphicx}%
\usepackage{amssymb}%
\usepackage{bm}%
\usepackage{epsfig}%
\usepackage{amsmath}%
\usepackage{bbm}%
\usepackage{makeidx}%
\usepackage{wrapfig}%
\usepackage{float}%
\usepackage{graphicx}%
\usepackage{dcolumn}%
\usepackage{bm}%
\usepackage{amsmath}%
\usepackage{amssymb}%
\usepackage{epsfig}%
\usepackage{natbib}%

\begin{document}

\title{The zero-temperature random-field Ising model on a bi-layered Bethe lattice}

\author{Thomas~P.~Handford}
\affiliation{Department of Chemistry, University of Cambridge, Cambridge, UK}
\author{Francisco~J.~P{\'e}rez-Reche}
\affiliation{Institute for Complex Systems and Mathematical Biology, SUPA, University of Aberdeen, Aberdeen, UK}
\author{Sergei~N.~Taraskin}
\affiliation{St. Catharine's College and Department of Chemistry,
University of Cambridge, Cambridge, UK}

\begin{abstract}
The zero-temperature random-field Ising model is solved analytically for magnetisation {\it vs} external field for a bi-layered Bethe lattice.
The mechanisms of infinite avalanches which are observed for small values of disorder are established. 
The effects of variable inter-layer interaction strengths on infinite avalanches are investigated.
The spin-field correlation length is calculated and its critical behaviour is discussed.
Direct Monte-Carlo simulations of spin-flip dynamics are shown to support the analytical findings.
We find, paradoxically, that a reduction of the inter-layer bond strength can cause a phase transition 
from a regime with continuous magnetisation reversal to a regime where magnetisation exhibits a discontinuity associated with an infinite avalanche.
This effect is understood in terms of the proposed mechanisms for the infinite avalanche.
\end{abstract}


\maketitle

\section{Introduction}

The non-equilibrium zero-temperature random-field Ising model (zt-RFIM) is a model for crackling noise in disordered systems such as ferromagnets~\cite{Sethna1993}, martensites~\cite{Rosinberg_Vives_Review2010}, condensation in porous media~\cite{Kierlik1998} and fracture of materials~\cite{Pradhan2010}.
Disorder in the zt-RFIM refers to quenched local random magnetic fields applied to each Ising spin in the system.
For this model, the hysteresis response of the magnetisation to changing external field is well established for spin systems with certain underlying topologies, including a complete graph, by means of the mean-field approximation~\cite{Sethna1993}, Bethe lattice~\cite{Dhar1997} and hyper-cubic lattices~\cite{Perkovic1999}. 
In the mean-field, for sufficiently low degree of disorder, 
the mean magnetisation exhibits a jump discontinuity as a function of external field.
This jump is reproduced on the Bethe lattice with coordination number $q>3$~\cite{Dhar1997}, and for hyper-cubic lattices in dimensions $d\ge 2$~\cite{Perkovic1999,Spasojevic_PRL2011}.
The reason by which the infinite avalanche in a Bethe lattice occurs for $q>3$ but not for $q\le 3$ has recently been understood in terms of the number of free paths available for the avalanche to propagate through the system~\cite{Handford2013Mech}.
In order for an infinite avalanche to occur, the flip of one neighbour of a spin has to increase the probability that a subsequent avalanche will branch when it arrives at that spin, i.e.
the probability that the spin flips and causes more than one of its neighbours to flip.
If this probability does increase then, after a certain external field is reached,
a single avalanche can grow exponentially by interacting with a large number of pre-flipped spins which flipped during previous finite avalanches.
This cannot occur in a $q=3$ lattice, as the flip of a single neighbour prevents branching occurring at that spin at a later time, due to the low coordination number and consequent blocking effect. 
The simplest way to avoid blocking is to increase the coordination number of the Bethe lattice to $q>3$, where, indeed, infinite avalanches occur.
However, it is an interesting question whether increasing the coordination number but keeping the global topology of the $q=3$ Bethe lattice can avoid the blocking effect or not.
In this paper, we address this question by considering the zt-RFIM on a bi-layered $q$-coordinated Bethe lattice,
in which every spin is replaced by a pair of spins connected to each other~\cite{Lyra1992,Hu1997}.
The interaction strength between spins within each layer is chosen to be the same but interactions between spins in the two layers is a variable parameter.

Our main finding is that an infinite avalanche can occur for a bi-layered Bethe lattice, and we suggest a possible mechanism for avoiding blocking effects.
Variation in the inter-layer interaction strength is found to result in non-monotonic behaviour for both the critical degree of disorder below which the discontinuity occurs and for the size of the jump in magnetisation when it does occur.
In particular, there are regions of the space disorder-interaction in which a decreasing strength of interaction between layers leads to a continuous (second order) transition in the size of the jump in magnetisation from zero to finite and \emph{increasing} values. 
Such a behaviour, i.e. bigger avalanches caused by smaller inter-layer interaction, which might be considered counter-intuitive, is consistent with the suggested mechanisms of infinite avalanches.

Technically, the zt-RFIM in the bi-layered Bethe lattice has been analysed by calculating the mean magnetisation as a function of the external field by means of a method employing a stochastic tensor formalism. 
This is a generalisation of the method used in~\cite{Sabhapandit2004} for quasi-1d lattices. 
The method has allowed exact solutions for mean magnetisation and correlation length as a function of external field in a bi-layered Bethe lattice to be obtained.

The paper is structured as follows.
The model is formulated in Sec.~\ref{sec:Model} and the solution for magnetisation is presented in Sec.~\ref{sec:Bilayer}.
Results for the magnetisation {\it vs} external field and the phase diagram are given in Sec.~\ref{sec:PhaseDiagram}.
A formula for the correlation length is derived in Sec.~\ref{sec:Correlations}.
In Sec.~\ref{sec:Mechanism} we explain the mechanism of the infinite avalanche on a bi-layered lattice.
The effect of altering the inter-layer interaction strength is analysed in Sec.~\ref{sec:Interlayer} and the conclusions are in Sec.~\ref{sec:Conclusions}.

\section{Model}\label{sec:Model}

The non-equilibrium zt-RFIM describes a set of $N$ spins $\{s_i\}$ placed on the nodes of a network of certain topology.
The spins interact with each other, with local random fields, $h_i$, and with an external field, $H$, according to the following Hamiltonian,
\begin{equation}
{\cal H} = -\sum_{\langle ij\rangle}J_{ij}s_is_j-H\sum_is_i-\sum_i{h_is_i}~,\label{eq:Hamiltonian}
\end{equation}
where $J_{ij}>0$ are the strengths of ferromagnetic interactions between neighbouring spins $i$ and $j$.
The local random fields are quenched and independent, and distributed according to a probability density function (p.d.f.) $\rho(h_i)$.

If the external field increases monotonically and adiabatically from $-\infty$ (when all spins are down, $s_i=-1$), 
and the dynamical rules for relaxation are such that spins flip one at a time
then the system will pass through a sequence of states which are stable to single spin flips.
Each of these states is fully determined by the values of $h_i$ and $H$, i.e. they do not depend on the order in which spins flip (the system obeys the Abelian property)~\cite{Sethna1993}. 

During the process of relaxation, the spins individually align themselves with their local fields, $f_i$, given by, 
\begin{equation}
f_i=H+h_i+\sum_{j/i}J_{ij}s_j~,\label{eq:localField}
\end{equation}
where the sum is taken over all neighbours, $j$, of spin $i$.
The main quantity of interest in this paper is the mean magnetisation of the system,
\begin{equation}
\langle m \rangle = N^{-1}\sum_is_i~.\label{eq:magnetisation}
\end{equation}
and its dependence on external field, i.e. the hysteresis loop.

\section{Magnetisation of bi-layered Bethe lattice}\label{sec:Bilayer}

\begin{figure*}
\includegraphics[width=10.0cm]{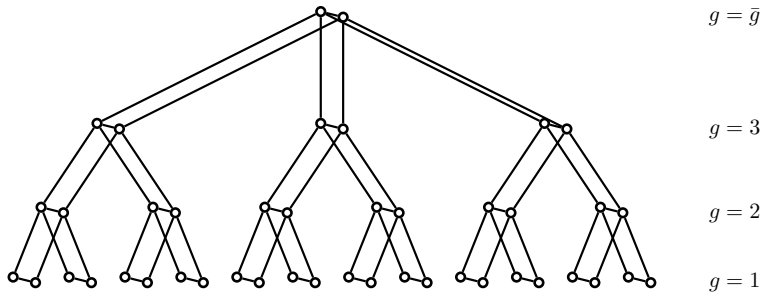}
\caption{Diagram of a bi-layered Cayley tree as considered in this paper. In this example, ${\bar g}=4$, $q=3$ and $z=4$.
\label{fig:bilayeredCayley}}
\end{figure*}

In this section, we derive an iterative procedure to calculate the mean magnetisation of a bi-layered Bethe lattice in the zt-RFIM.
A bi-layered Bethe lattice is defined as two copies of a single-layered Bethe lattice with coordination number $q$, where a link is added to connect each node in the first lattice to its copy in the second lattice (see Fig.~\ref{fig:bilayeredCayley}). 
The interaction of neighbouring spins in the same layer is characterised by $J_{ij}=J$, while interactions between spins in different layers is described by $J_{ij}=J^\prime$, which does not necessarily coincide with $J$.
The presence of inter-layer links results in the coordination number of each spin in a bi-layered Bethe lattice being $z=q+1$. 
The topology of the bi-layered lattice is, however, significantly different from that of a Bethe lattice with coordination number $q+1$. 
For instance, there exist local loops in the bi-layered lattice which are absent, by definition, in a single-layered Bethe lattice.

Such a bi-layered lattice can be considered as a single-layered lattice where each node is replaced by a pair of spins which are connected to each other and to the neighbouring spin pairs. 
The spin state, ${\bm s}_X$, of an arbitrary pair of spins, $X$, in the lattice can be characterised by a $2$-dimensional state vector ${\bm s}_X=(s_{X_1},s_{X_2})$, where each component is a discrete variable taking values $\pm 1$ and describing the states, $s_{X_1}$ and $s_{X_2}$, of the spins in layers $\ell=1$ and $\ell=2$, respectively.
Such a pair can be in four possible states, i.e. ${\bm s}_X=(s_{X_1},s_{X_2})={\cal A}$, ${\cal B}_1$, ${\cal B}_2$ or ${\cal C}$, where ${\cal A} = (-1,-1)$, ${\cal B}_1=(+1,-1)$, ${\cal B}_2=(-1,+1)$ and ${\cal C}=(+1,+1)$.

We consider bi-layered Bethe lattice as the infinite limit, ${\bar g}\to\infty$, of a bi-layered Cayley tree consisting of ${\bar g}$ generations of spins. 
In this bi-layered Cayley tree, spins in generation $g=1$ are linked to their copy in the other layer, and to a single spin in generation $g=2$ (see Fig.~\ref{fig:bilayeredCayley}). 
Spins in generation $1<g<{\bar g}$ are each linked to their copy, a single spin in generation $g+1$ and $q-1$ spins in generation $g-1$. 
The $2$ root spins, $R$, in generation $g={\bar g}$ are linked to $q$ spins in generation ${\bar g}-1$ and to each other.
In the limit ${\bar g}\to\infty$, the mean magnetisation of the bi-layered Bethe lattice, equal to the mean magnetisation of the one of root spins, $\langle m_{\text{R}_1}\rangle$, is given by,
\begin{equation}
\langle m\rangle=2P(s_{\text{R}_1}=1)-1~,\label{eq:magnetisationBiLayer}
\end{equation}
where $P\left(s_{\text{R}_1}=1\right)$ is the probability that the root spin in layer $1$ is in the up-state ($P\left(s_{\text{R}_1}=1\right)=P\left(s_{\text{R}_2}=1\right)$).

\begin{figure}
\includegraphics[width = 8cm]{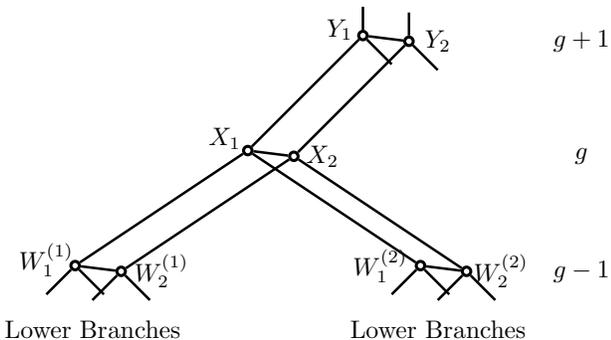}
\caption{A pair of spins at generation $g$ and its neighbours in generations $g-1$ and $g+1$. 
\label{fig:twoPairs}}
\end{figure}

The probability $P(s_{R_1}=1)$ can be calculated for an arbitrary value of $H$ using an iterative procedure. 
In particular, we apply the method to calculate the value of $P(s_{R_1}=1)$ for relaxation from an
initial state when all the spins are down. 
This gives a dependence of $P(s_{R_1}=1)$ on $H$ which corresponds to the lowest branch of the magnetisation hysteresis loop. 
For any given $H$, the calculation proceeds as follows.
Before the first step, all spins are artificially held in the down-state.
At step $1$, the spins in generation $g=1$ are allowed to flip if their local field is positive.
After this, at step $2$, spins in all generations $g\le 2$ are allowed to flip if their local field is positive.
Such a process continues progressively until all spins in the whole tree with a positive local fields have flipped at step ${\bar g}$.
At each step,
avalanches of spin flip events will pass through the part of the system in which spins are allowed to flip 
eventually leading to the possibility that an avalanche will reach one of the root spins, $R_\ell$ ($\ell=1,2$), causing it to flip. 
For concreteness, we choose to analyse the magnetisation of root spin in layer $1$, $R_1$. 

In order to find the probability that the root spin flips, we should account for all possible routes that avalanches can take to the root spin, which can be done by an inductive method, being the standard approach for the Bethe lattice~\cite{Baxter_Book}. 
At each step of the inductive process we derive an expression for the state function,
${\bm s}_{X}={\bm s}_{X}({\bm s}_Y)$, 
which describes how the state of pair $X$ in generation $g$ depends on
the state of the pair $Y$ in the generation above (see Fig.~\ref{fig:twoPairs}).
The form of this function depends on similar state functions, ${\bm s}_{W^{(b)}}={\bm s}_{W^{(b)}}({\bm s}_X)$, describing the state of the neighbouring pairs, 
$W^{(b)}$ ($b=1,\ldots,q-1$), in generation $g-1$ as a function of the state of the pair $X$.
Thus, by finding the relationship between ${\bm s}_{X}({\bm s}_Y)$ and ${\bm s}_{W^{(b)}}({\bm s}_X)$, an inductive process can be established. 
By continuing the inductive procedure to the spins neighbouring the
root pair (at generation ${\bar g}-1$), all avalanches propagating into the
root pair can be described, allowing the final state, $s_{R_1}$, of
the root spin to be determined.

The state function ${\bm s}_{X}({\bm s}_Y)$ depends on the random fields at spin pair ${\bm s}_X$ and in the branches below it and is a discrete random function which has $4$ possible states for each of the $4$ states of ${\bm s}_Y$, and thus has $4^4=256$ possible functional forms.
It is convenient, however, to categorise each functional form into one of $5$ categories, labelled $C_i$ ($1\le i\le 5$). 
Categories specify the possible changes of a pair $X$ at generation $g$ from its initial state, ${\bm s}_X={\cal A}$, induced by changes of state of pairs in neighbouring generations, $g-1$ and $g+1$. 
Table \ref{Table_1} summarises the categories and a detailed definitions are given in App.~\ref{sec:Categories}. Categories $C_1$ and $C_2$ include cases in which ${\bm s}_X$ can only change as a consequence of a change of the pair ${\bm s}_Y$ in the generation above. 
After a change ${\cal A} \to {\cal B}_1$ of state of the pair $Y$, category $C_1$ accounts for the case in which ${\bm s}_X$ does not change, or changes to a stable state ${\cal B}_1$ (see Fig.~\ref{fig:categories1} in App.~\ref{sec:Categories}). 
In contrast, category $C_2$ describes the case in which ${\bm s}_X={\cal B}_1$ is not stable and an \emph{in-out} avalanche occurs which leads to the state ${\bm s}_X={\cal C}$ (Fig.~\ref{fig:categories2}). 
The term in-out avalanche refers to the fact that such a change of the state ${\bm s}_X$ as a result of the flip of $s_{Y_1}$ (avalanche going in to the branch) can increase the local field at spin $s_{Y_2}$ (avalanche going out of the branch).
Since $s_{Y_2}$ is in the down-state, an in-out avalanche potentially causes it to flip and an avalanche to propagate back along the same branch, but in a different layer.
Categories $C_3$, $C_4$ and $C_5$ describe cases in which ${\bm s}_X$ changes before ${\bm s}_Y$ does (\emph{spontaneous} flips; see Figs.~\ref{fig:categories3}-\ref{fig:categories5} in
App.~\ref{sec:Categories}). 
Here, spontaneous flip refers to the fact that these spin flips were not triggered by an avalanche coming from the upper generation.
The change of ${\bm s}_X$ may occur because the state ${\bm s}_X={\cal A}$ is not stable at the external field, $H$, irrespective of the state of its neighbours or may be induced by spin flips in lower generations. 
Categories $C_3$ and $C_4$ account for changes of state that involve the flip of a single member of the pair ${\bm s}_X$ and $C_5$ describes events in which both spins of the pair $X$ flip.

\begin{table}
\caption{\label{Table_1} Summary of the categories $C_i$ of the changes of spin pair ${\bm s}_X$ given the state ${\bm s}_Y$ in the generation above. 
The third column describes the sequence of states taken by spin pair ${\bm s}_X$ given the sequence of states taken by spin pair ${\bm s}_Y$ (second column).}
\begin{tabular}{l!{ }c!{ }c}
\hline
Category & ${\bm s}_Y$ & ${\bm s}_X$\\
\hline
$C_1$ & ${\cal A}$ & ${\cal A}$ \\
      & ${\cal A} \to {\cal B}_1$ & ${\cal A} \to {\cal A}$ or ${\cal A} \to {\cal B}_1$\\
$C_2$ & ${\cal A}$ & ${\cal A}$ \\
      & ${\cal A} \to {\cal B}_1$ & ${\cal A} \to {\cal B}_1 \to {\cal C}$\\
$C_3$ & ${\cal A}$ & ${\cal A} \to {\cal B}_1$ \\
$C_4$ & ${\cal A}$ & ${\cal A} \to {\cal B}_2$ \\
$C_5$ & ${\cal A}$ & ${\cal A} \to {\cal B}_1 \to {\cal C}$ or ${\cal A} \to {\cal B}_2 \to {\cal C}$ or ${\cal A} \to {\cal C}$\\
\hline
\end{tabular}
\end{table}

The behaviour of spin pair ${\bm s}_X$ with respect to avalanches (i.e. the category of ${\bm s}_X({\bm s}_Y)$) is determined entirely by the local fields $f_{X_\ell}$, which in turn depend on the random fields, the external field and the interaction terms with all neighbouring spins. 
The interactions with spin pairs in the generation below depend on the state vectors of those spins, which are determined by the categories, $C_i$, either of their state function, ${\bm s}_{W^{(b)}}({\bm s}_X)$ or of the state function ${\bm s}_{W^{(b)}}^\prime({\bm s}_X^\prime)$.
Here, ${\bm s}_{W^{(b)}}^\prime({\bm s}_X^\prime)$ is the state function that spin pair $W^{(b)}$ would have if all spins in layer $\ell=1$ were swapped with their copies in layer $\ell=2$, i.e. ${s}_{W^{(b)}_1}^\prime({\bm s}_X^\prime)={s}_{W^{(b)}_2}({\bm s}_X)$ and ${s}_{W^{(b)}_2}^\prime({\bm s}_X^\prime)={s}_{W^{(b)}_1}({\bm s}_X)$ where ${s}_{X_1}^\prime={s}_{X_2}$ and ${s}_{X_2}^\prime={s}_{X_1}$. 

The function ${\bm s}_{X}({\bm s}_Y)$ falls into a category $C_i$ randomly with probability $P(C_i,g)$, which can be represented, for generation $g$, as a $5$-dimensional vector, ${\bm P}^{(g)}$, with components ${P}^{(g)}_{i} = P(C_i,g)$. 
Similarly, the functions ${\bm s}_{W^{(b)}}({\bm s}_X)$, or equivalently, ${\bm s}^\prime_{W^{(b)}}({\bm s}^\prime_X)$, fall into category $C_i$ with a probability $P(C_i,g-1)$, which is represented by the same $5$-dimensional vector, ${\bm P}^{(g-1)}$.
In App.~\ref{sec:SelfConsistent}, a recursion relation is derived to describe ${\bm P}^{(g)}$ in terms of ${\bm P}^{(g-1)}$, which takes the form of a tensor relationship,
\begin{eqnarray}
{P}^{(g)}_{i} = A_{ijk \dots l}P^{(g-1)}_{j}P^{(g-1)}_{k}\dots P^{(g-1)}_{l}~,\label{eq:tensorRecursion}
\end{eqnarray}
where $1\le i,j, \dots, l\le 5$, summation is assumed over repeated indexes and $\bm A$ is a tensor of rank $q$ given by,
\begin{subequations}\label{eq:AMatrix}
\begin{eqnarray}
&&A_{1jk\dots l}=(1-p_{(n_3+n_5),0})(1-p_{(n_4+n_5+1),0})\nonumber\\
&&+(1-p_{(n_2+n_3+n_5),1})(p_{(n_4+n_5+1),0}-p_{(n_4+n_5),0})
\end{eqnarray}
\begin{eqnarray}
A_{2jk\dots l} &=& (p_{(n_2+n_3+n_5),1}-p_{(n_3+n_5),0})\nonumber\\
&\times&(p_{(n_4+n_5+1),0}-p_{(n_4+n_5),0})
\end{eqnarray}
\begin{eqnarray}
A_{3jk\dots l} &=& (1-p_{(n_2+n_3+n_5),1})p_{(n_4+n_5),0}\label{eq:AMatrix3}
\end{eqnarray}
\begin{eqnarray}
A_{4jk\dots l} &=& p_{(n_3+n_5),0}(1-p_{(n_2+n_4+n_5),1})
\end{eqnarray}
\begin{eqnarray}
A_{5jk\dots l} &=& p_{(n_3+n_5),0}p_{(n_2+n_4+n_5),1}\nonumber\\
&+&(p_{(n_2+n_3+n_5),1}-p_{(n_3+n_5),0})p_{(n_4+n_5),0}~.
\end{eqnarray}
\end{subequations}
Here, the sub-scripts $n_t$ take integer values in $[0,q-1]$ given by the discrete function,
\begin{equation}
n_t(j,k, \dots, l)=\delta_{jt}+\delta_{kt}+\dots+\delta_{lt}~,\label{eq:numInCat}
\end{equation}
where $\delta_{ij}$ is the Kronecker-delta. 
The function $n_t(j,k, \dots, l)$ gives the number of the subscripts $j,k, \dots, l$ which are equal to $t$ ($0\le t\le 5$), with $0\le n_t \le q-1$. 
In Eqs.~\eqref{eq:AMatrix}, the values of $n_t$ specify the number of the state functions of spin pairs in the generation below which are of a certain category, $C_t$. 
For instance, in the case of $q=4$, the value of $A_{3521}$ can be calculated by setting $n_1=1$, $n_2=1$, $n_3=0$, $n_4=0$ and $n_5=1$ in Eq.~\eqref{eq:AMatrix3}, resulting in, $A_{3521}=(1-p_{2,1})p_{1,0}$. 
The values of $p_{n,n^\prime}$ ($0\le n\le q$ and $n^\prime=0,1$) are the probabilities of finding the local fields in certain intervals and are given by, 
\begin{equation}
p_{n,n^\prime}=\int_{h_i=-H-(2n-q)J-(2n^\prime-1)J^\prime}^\infty\rho(h_i)\text{d}h_i~.
\end{equation}

In Eqs.~\eqref{eq:AMatrix}, the first index ($i$) of $A_{ijk\dots l}$ refers to the category of state function ${\bm s}_X({\bm s}_Y)$, while the other indexes $j,k, \dots, l$ refer to the categories of the $q-1$ state functions, ${\bm s}_{W^{(b)}}({\bm s}_{X})$, of spin pairs in the generation below.
The tensor $\bm A$ is symmetric in all indexes except the first, $A_{ijk\dots l}=A_{i(jk\dots l)}$.
It can be described as a stochastic tensor, i.e. $\sum_iA_{ijk\dots l}=1$, and for the case of $q=2$ it is equivalent to the stochastic matrix given by Eq.~(4) in Ref.~\cite{Sabhapandit2004}. 

The probability that the root spin flips, $P(s_{R_1}=+1)$, can be calculated by a similar method to that used to obtain Eq.~\eqref{eq:tensorRecursion} 
(see App.~\ref{sec:rootSite}),
resulting in the following tensor relationship,
\begin{eqnarray}
P(s_{R_1}=1)&=&B_{ij\dots k}{P}^{({\bar g}-1)}_{i}{P}^{({\bar g}-1)}_{j}\dots {P}^{({\bar g}-1)}_{k}~,\label{eq:rootBilayerM}
\end{eqnarray}
where $\bm B$ is a tensor of rank $q$ given by,
\begin{eqnarray}
B_{ij\dots k} &=& p_{(m_3+m_5),0}+p_{(m_4+m_5),0}\nonumber\\&\times&(p_{(m_2+m_3+m_5),1}-p_{(m_3+m_5),0})
\end{eqnarray}
with the indexes $1\le i,j,\dots,k\le 5$ referring to the possible categories of the spin pairs in generation ${\bar g}-1$.
The values of $m_t$ ($0\le m_t\le q$) are given by the discrete function $m_t(i,j \dots k)=\delta_{it}+\delta_{jt}+\dots+\delta_{kt}$ which counts the number of spin pairs in the generation ${\bar g}-1$ that belong to a category $t$.

In the limit of a large number of generations, ${\bar g}\to\infty$, the vector ${\bm P}^{({\bar g})}$ tends to a limit, ${\bm P}^*=\lim_{h\to\infty}{\bm P}^{({\bar g})}$, given by,
\begin{eqnarray}
{P}^*_{i} = A_{ijk \dots l}P^*_{j}P^*_{k}\dots P^*_{l}~. \label{eq:recursiveBilayer}
\end{eqnarray}
This is a self-consistent equation which can be solved numerically for ${\bm P}^*$ (with the constraints $0\le P^*_i\le 1$ and $\sum_iP^*_i=1$).
Finally, the mean magnetisation, $\langle m\rangle = 2P(s_{R_1}=+1)-1$, of the bi-layered Bethe lattice can be found by substituting ${\bm P}^*$ into Eq.~\eqref{eq:rootBilayerM}.

\section{Phase Diagram}\label{sec:PhaseDiagram}

\begin{figure}
\includegraphics[width=8.0cm]{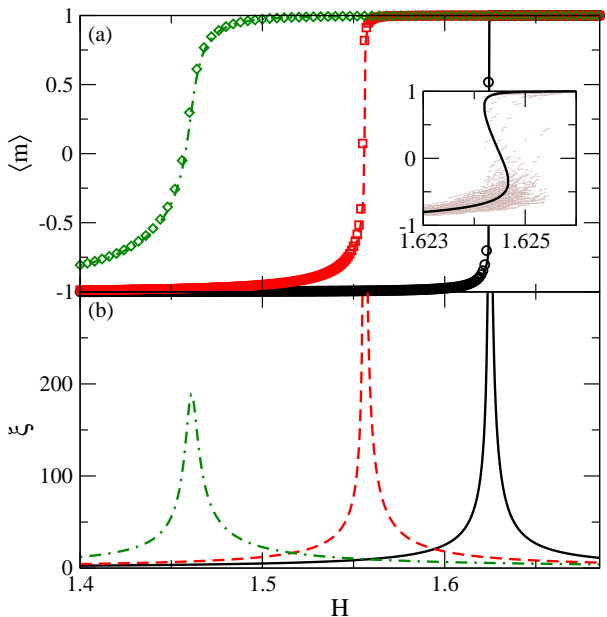}
\caption{Panel (a): The mean magnetisation $\langle m\rangle$ {\it vs} $H$ for a bi-layered Bethe lattice with $q=3$, $\Delta/J=1.0$ (Dot-dashed line) $\Delta/J=\Delta_{\text{c}}/J=0.8171$ (Dashed Line) and $\Delta/J=0.69$ (Solid Line). 
The lines show the solution given by Eqs.~\eqref{eq:AMatrix}-\eqref{eq:recursiveBilayer} while the symbols are the results of numerical simulations on a system of $N=2\times 10^7$ spins averaged across $10^2$ realisations of disorder.  
The inset magnifies the small multivalued region for the case $\Delta/J=0.69$. 
Individual data points in the inset represent the data before averaging over realisations of quenched disorder in random fields and network structure of bi-layered $3$-regular graph.
Panel (b): The correlation length $\xi$ as a function of $H$ obtained from the numerical solution of exact equations derived in Sec.~\ref{sec:Correlations}. Same line styles as in panel (a).
\label{fig:solutionBiLayer}}
\end{figure}

\begin{figure}
\includegraphics[width=8.0cm]{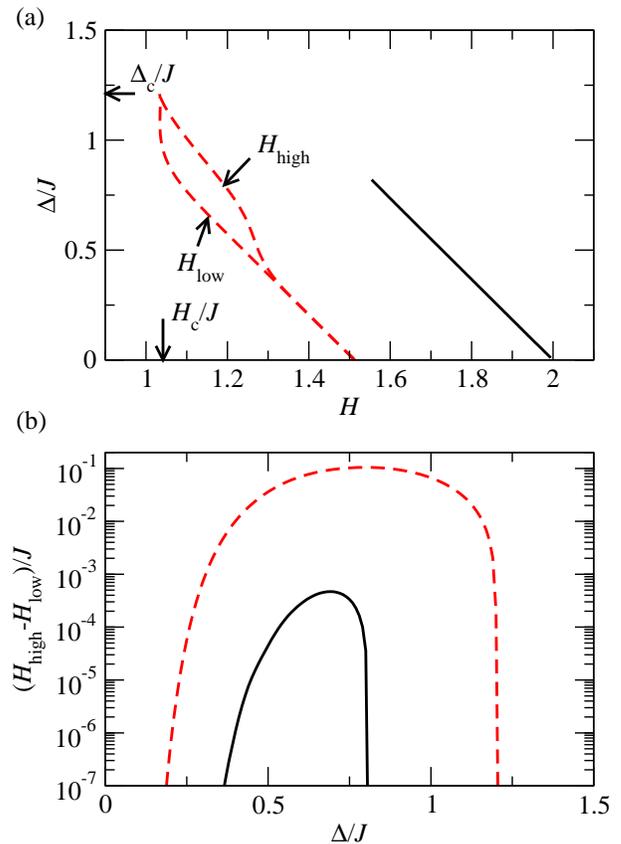}
\caption{Size of multivalued region varying with disorder for an interlayer link strength of $J'/J=1.0$ (solid line) and $J'/J=0.5122$ (dashed). 
The difference between $H_{\text{high}}$ and $H_{\text{low}}$ for $J'/J=1.0$ cannot be appreciated at the scale of the graph.
Panel (b): The inset shows the width, $H_{\text{high}}-H_{\text{low}}$, of the multivalued region as a function of disorder $\Delta$. 
\label{fig:phaseDiagram}}
\end{figure}

In this section, we analyse the behaviour of magnetisation as a function of external field for different values of disorder, in the case where the inter-layer and in-layer interaction strengths coincide, $J=J^\prime$.
The results of numerical solution for $\langle m\rangle(H)$ of the self-consistent equations Eqs.~\eqref{eq:AMatrix}-\eqref{eq:recursiveBilayer} is shown in Fig.~\ref{fig:solutionBiLayer}(a) for normally distributed random fields with mean equal to zero and variance $\Delta^2$, i.e. $h_i\sim N(0,\Delta^2)$. 
Different lines, describing the lower branch (increasing $H$) of the hysteresis loop, correspond to different degrees of disorder, represented by the value of $\Delta/J$.
As can be seen in the inset of Fig.~\ref{fig:solutionBiLayer}(a), there is a region in the parameter space of $H$ and $\Delta$ in which the solution for mean magnetisation is multivalued.
The phase diagram displayed in Fig.~\ref{fig:phaseDiagram}(a) shows the extent of this region.

The region in which $\langle m\rangle(H)$ is multivalued exists when the degree of disorder is less than the critical value, $\Delta <\Delta_c$, in a finite range of external fields $H_{\text{low}}(\Delta)<H<H_{\text{high}}(\Delta)$. 
At the critical value of disorder, $\Delta_{\text{c}}/J \simeq 0.8171$, the width of the region shrinks to $0$, i.e. $H_{\text{low}}(\Delta_{\text{c}})/J=H_{\text{high}}(\Delta_{\text{c}})/J=H_c/J=1.556$. 
The width of the region, $H_{\text{high}}(\Delta_{\text{c}})-H_{\text{low}}(\Delta_{\text{c}})$, has its maximum value for $\Delta/J\simeq 0.69$ (see Fig.~\ref{fig:phaseDiagram}(b) and inset of Fig.~\ref{fig:solutionBiLayer}(a)) and decreases rapidly for $\Delta/J<0.69$. 
In the interior of the region, there are 
three solutions to Eqs.~\eqref{eq:rootBilayerM} and~\eqref{eq:recursiveBilayer}, $\langle m_1(H)\rangle<\langle m_2(H)\rangle<\langle m_3(H)\rangle$ (see inset in Fig.~\ref{fig:solutionBiLayer}(a)), two of which merge at the boundaries of the region, i.e. $\langle m_2(H_{\text{low}})\rangle=\langle m_3(H_{\text{low}})\rangle$ at the left boundary and $\langle m_1(H_{\text{high}})\rangle=\langle m_2(H_{\text{high}})\rangle$ at the right boundary. 
Outside of the region, the solution is single valued, taking the values $\langle m_1\rangle(H)$ for $H<H_{\text{low}}$ and $\langle m_3(H)\rangle$ for $H>H_{\text{high}}$. 
For increasing external field and fixed degree of disorder, $\Delta<\Delta_{\text{c}}$, 
the magnetisation will follow the path of the solution with the lowest magnetisation, namely, $\langle m\rangle = \langle m_1(H)\rangle$ for $H\le H_{\text{high}}$ and $\langle m\rangle = \langle m_3(H)\rangle$ for $H>H_{\text{high}}$, thus exhibiting a jump of size $\delta m(\Delta)=\langle m_3(H_{\text{high}})\rangle-\langle m_1(H_{\text{high}})\rangle$ as $H$ passes $H_{\text{high}}$. 

Mathematically, the origin of the discontinuity in magnetisation can be understood in terms of a diverging derivative of $P^*_i$ with respect to $H$.
Differentiation of Eq.~\eqref{eq:recursiveBilayer} leads to,
\begin{equation}
\frac{\text{d}P^*_i}{\text{d}H}=\left[(\bm I-\bm D)^{-1}\right]_{ii^\prime}\frac{\partial A_{i^\prime jk\dots l}}{\partial H}P^*_jP^*_k\dots P^*_l~,\label{eq:BiLayerCondition}
\end{equation}
where the matrix $\bm D$ is defined as,
\begin{equation}
D_{ij}=(q-1)A_{ijk\dots l}P^*_k\dots P^*_l~.\label{eq:Dmatrix}
\end{equation}
The non-negative matrix $\bm D$ always has one eigenvalue equal to $q-1$, which ensures that $\sum_i\text{d}P^*_i/\text{d}H=0$ (so the probabilities $P^*_i$ always sum to unity). 
At the discontinuity in the magnetisation, one of the other eigenvalues, $\lambda$, passes through $1$, causing the matrix $({\bm I}-{\bm D})$ to become singular.

The scaling exponent describing $\langle m\rangle$ as a function of external field around the critical point $(H_c,\Delta_c)$ can be computed by solving numerically the exact equations giving $\langle m\rangle$, and is found to be equal to the mean-field value, i.e. $\delta=3$ in $\langle m\rangle\propto|H-H_C|^{1/\delta}$ (at fixed $\Delta=\Delta_c$)~\cite{Sethna1993}.
Similarly, the size of the jump as a function of $\Delta$ scales with the mean-field exponents,
i.e. $\beta=1/2$ for $\delta m(\Delta)\propto|\Delta-\Delta_C|^{\beta}$. 

In order to test the analytical results, we have performed numerical simulations on a finite size ($N=2\times 10^7$) bi-layered $3$-regular graph, which is a finite-size approximation of a Bethe lattice~\cite{Dhar1997,NewmanNetworks}. 
Such a graph is formed of $N/2$ pairs of spins in layers $1$ and $2$, out of which two pairs are selected at random to be connected together (by introducing a link between the members of each pair belonging to the same layer only) repeatedly until all pairs are connected to exactly $3$ other pairs.
The random fields at each spin were chosen from a normal distribution, and the external field $H$ was set initially to $-\infty$, forcing all the spins into the down-state.
The field was then increased to the value required to make the spin with highest local field flip.
Avalanches of spin flips were then allowed to pass through the lattice until all spins were stable.
When avalanches had ceased, $H$ was increased until another avalanche started.
This process was repeated until all spins had flipped.
The magnetisation was measured as a function of external field, and the resulting hysteresis loops are shown by symbols in Fig.~\ref{fig:solutionBiLayer}(a) and the points in its inset.
As follows from this figure, the numerical results support the analytical findings.

\section{Correlation Length}\label{sec:Correlations}

Using an extension of the method discussed in Sec.~\ref{sec:Bilayer}, the correlation length $\xi$ for field-spin correlations in the bi-layered Bethe lattice can be derived.
Indeed, if a small test field, $\delta h$, is applied to a spin in generation ${\bar g}-r$ ($1\ll r\ll {\bar g}$), it is expected that the magnetisation of one of the central spins will change by an amount $\langle\delta m\rangle\sim \left[N_q(r)\right]^{-1}\exp(-r/\xi)\delta h$, where $\xi$ gives a correlation length~\cite{ChristensenBOOK,Handford2012} 
and the factor $N_q(r)\propto (q-1)^r$ is the number of spins at a distance $r$ from the root.
In order to calculate $\xi$, we assume that the test field has been applied to all of the spins in generation ${\bar g}-r$ and that the resulting change in magnetisation, $\langle\delta m^\prime\rangle$, is related to $\langle\delta m\rangle$ according to $\langle\delta m^\prime \rangle \propto \langle\delta m\rangle N_q(r)$. 

The value of $\langle\delta m^\prime\rangle$ can be estimated as follows.
The test field applied to the spins in generation ${\bar g}-r$ will alter the value of ${\bm P}^{({\bar g}-r)}$ given by Eq.~\eqref{eq:tensorRecursion}, meaning that,
\begin{equation}
\bm P^{({\bar g}-r)}=\bm P^*+\delta\bm P^{({\bar g}-r)}~,
\end{equation}
where the vector $\delta\bm P^{({\bar g}-r)}$ has components,
\begin{equation}
\delta P^{({\bar g}-r)}_i\simeq\frac{\partial A_{ijk\ldots l}}{\partial H}P^*_jP^*_k\ldots P^*_l\delta h~.\label{eq:perturbr}
\end{equation}
As a result of the application of the test field, the values of $\bm P^{({\bar g}-r^\prime)}$ ($1<r^\prime<r$) also change by a small amount, i.e. $\bm P^{({\bar g}-r^\prime)}=\bm P^*+\delta\bm P^{({\bar g}-r^\prime)}$.
A recursive relation giving $\delta {\bm P}^{({\bar g}-r^\prime)}$ in terms of $\delta {\bm P}^{({\bar g}-r^\prime-1)}$ can be written using Eq.~\eqref{eq:tensorRecursion},
\begin{equation}
\delta P_i^{({\bar g}-r^\prime)}=D_{ij}\delta P^{({\bar g}-r^\prime-1)}_j
\end{equation}
so that $\delta\bm P^{({\bar g}-1)}=\bm D^{r-1}\delta\bm P^{({\bar g}-r)}$, and the value of $\langle m\rangle$ given by Eq.~\eqref{eq:rootBilayerM} changes by an amount,
\begin{equation}
\langle\delta m^\prime\rangle=\bm E^{\text{T}}\bm D^{r-1}\delta\bm P^{({\bar g}-r)}~,
\end{equation}
where the components of $\bm E$ are given by,
\begin{eqnarray}
E_{i}&=&\frac{\text{d}}{\text{d}P^*_i}B_{jk\ldots l}P^*_jP^*_k\ldots P^*_l
=qB_{ij\ldots l}P^*_j\ldots P^*_l~.\label{eq:CentralCond}
\end{eqnarray}

The values of $\bm D$, $\delta\bm P^{({\bar g}-r)}$ and $\bm E$ given by Eqs.~\eqref{eq:BiLayerCondition}, \eqref{eq:perturbr} and~\eqref{eq:CentralCond}, respectively, do not depend on $r$, meaning that $\langle\delta m^\prime\rangle\propto \lambda^r$, where $\lambda$ is the largest eigenvalue of $\bm D$, excluding the eigenvalue equal to $q-1$.
Equating $\langle\delta m\rangle\propto \left[N_q(r)\right]^{-1}\exp(-r/\xi)$ and $\langle\delta m\rangle\propto \left[N_q(r)\right]^{-1}\langle\delta m^\prime\rangle\propto \left[N_q(r)\right]^{-1}\lambda^{r}$, gives the correlation length as $\xi=-1/\ln(\lambda)$.
The correlation length is plotted in Fig.~\ref{fig:solutionBiLayer}(b) as a function of $H$ for values of $\Delta$ above, equal to and below $\Delta_{\text{c}}$, and is found to diverge at both the critical point and the magnetisation discontinuity. 
At the critical disorder, it diverges as $\xi(H,\Delta_{\text{c}})\propto |H-H_{\text{c}}|^{-\mu}$ with $\mu=2/3$. 
The divergence along the magnetisation discontinuity behaves as $\xi(H^*(\Delta),\Delta)\propto |\Delta-\Delta_{\text{c}}|^{-\nu}$, where $\nu=1$ and $H^*(\Delta)$ is a straight line passing through the critical point with the gradient of the curves $H_{\text{low}}(\Delta)$ and $H_{\text{high}}(\Delta)$ that are identical at $\Delta=\Delta_\text{c}$,
\begin{eqnarray}
\frac{\text{d}H^*}{\text{d}\Delta}
&=&\left.\frac{\text{d}H_{\text{high}}}{\text{d}\Delta}\right|_{\Delta_{\text{c}}}
=\left.\frac{\text{d}H_{\text{low}}}{\text{d}\Delta}\right|_{\Delta_{\text{cf}}}~.
\end{eqnarray}
The values of the critical exponents are identical to those found in a single-layered Bethe lattice~\cite{Handford2012}, indicating that the universality class has not been changed by the introduction of the second layer.

\section{Mechanism of Infinite Avalanche}\label{sec:Mechanism}

\begin{figure*}[t]
\includegraphics[width=14cm]{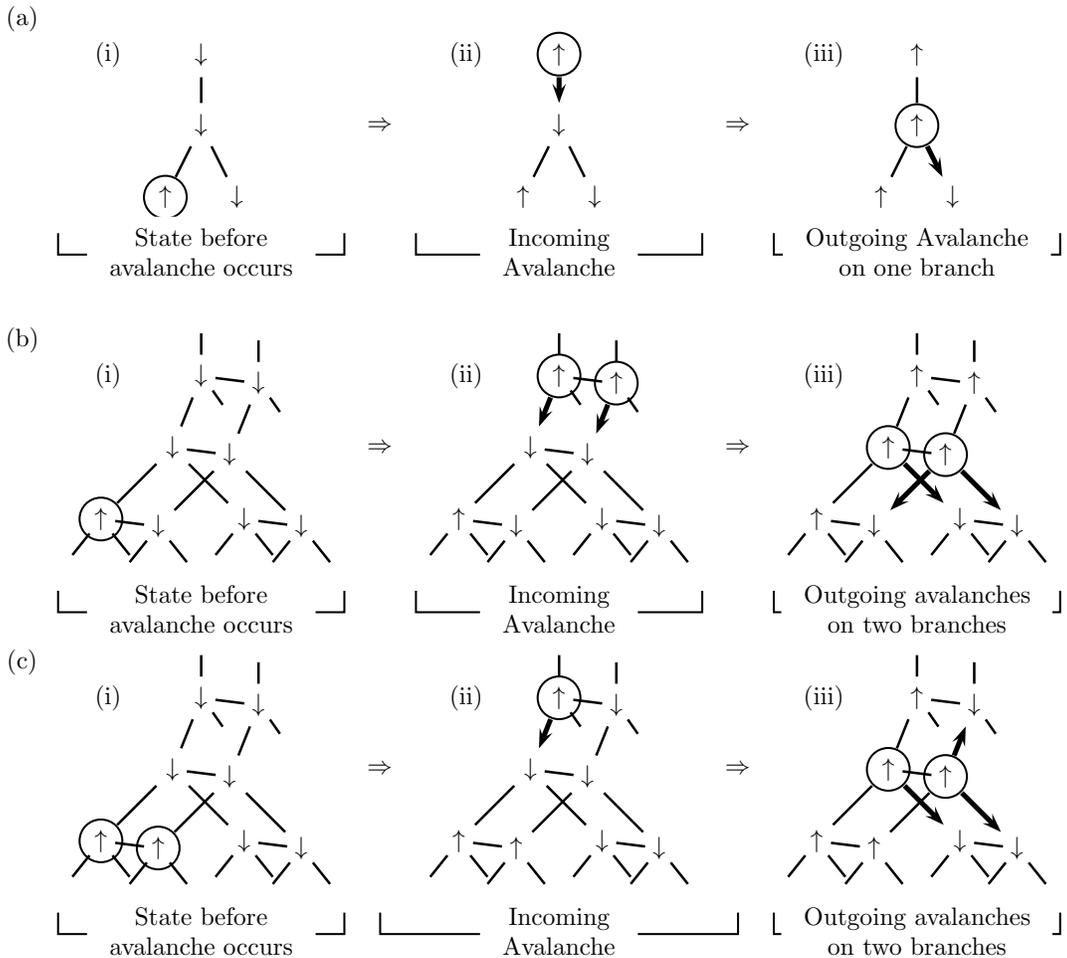}
\caption{\label{fig:BiLayerMechanisms} The interaction of an avalanche with pre-flipped spins in (a) a Bethe lattice of $q=3$, (b,c) a bi-layered Bethe lattice with $q=3$.
The circled spins in columns (i) and (ii) correspond to pre-flipped spins and spins which have flipped in a new approaching avalanche, respectively. 
In column (iii) the circles indicate that the central spins have flipped.
The bold arrows in columns (ii) and (iii) represent the incoming and outgoing avalanches with respect to the central pair, respectively.
}
\end{figure*}

It is known that an infinite avalanche does not take place in a $q=3$ single-layered Bethe lattice.
This is due to the fact that the infinite avalanche can only occur when avalanches branch at spins neighbouring pre-flipped spins (spins which flipped in previous avalanches). 
In a single-layered $q=3$ Bethe lattice, the pre-flipped spin blocks one of the out-going paths (see circled spin in Fig.~\ref{fig:BiLayerMechanisms}(a)(i)) meaning that a new avalanche (circled spin in Fig.~\ref{fig:BiLayerMechanisms}(a)(ii)) that reaches the central spin (bold arrow in Fig.~\ref{fig:BiLayerMechanisms}(a)(ii)) 
and causes it to flip (circled spin in Fig.~\ref{fig:BiLayerMechanisms}(a)(iii))
can only continue down one branch (bold arrow in Fig.~\ref{fig:BiLayerMechanisms}(a)(iii)), 
i.e. branching cannot occur at a spin neighbouring a pre-flipped spin.

Since there is a phase transition in the bi-layered $q=3$ lattice (see Fig.~\ref{fig:solutionBiLayer}), we can infer that blocking is avoided for such a system, i.e. branching must occur at a pair of spins neighbouring pre-flipped spins.
There are two simple mechanisms by which this can occur.
In the first mechanism, only one member of a pair of spins is pre-flipped (see circled spin in Fig.~\ref{fig:BiLayerMechanisms}(b)(i)), so that a new avalanche (circled spins in Fig.~\ref{fig:BiLayerMechanisms}(b)(ii)) which causes the central spins to flip (circled spins in Fig.~\ref{fig:BiLayerMechanisms}(b)(iii))
can continue along two branches (see the three bold arrows in Fig.~\ref{fig:BiLayerMechanisms}(b)(iii); the two parallel bold arrows correspond to an avalanche along the down-right branch in both layers and the third bold arrow shows an avalanche along the down-left branch in a single layer), i.e. the avalanche splits into two paths (two-way branching).
In the second mechanism, there are pre-flipped spins in both layers of one branch (circled spins in Fig.~\ref{fig:BiLayerMechanisms}(c)(i)) and a new avalanche causes a single spin neighbouring the central spins to flip (circled spin in Fig.~\ref{fig:BiLayerMechanisms}(c)(ii)).
This avalanche, in turn, causes the central spins to 
flip (circled spins in Fig.~\ref{fig:BiLayerMechanisms}(c)(iii)) and induces two-way branching (see the three bold arrows in Fig.~\ref{fig:BiLayerMechanisms}(c)(iii) along the upward branch and down-right branch).
In addition, two-way or even three-way branching can occur when one of the neighbouring pairs has one pre-flipped spin and the avalanche approaches on a single level. 

\section{The effect of inter-layer interaction strength}\label{sec:Interlayer}

Spin pairs with two spins in different states play an essential role in causing the infinite avalanche to occur
(see Sec.~\ref{sec:Mechanism}).
However, we have found numerically that such spin pairs are relatively rare for the case $J^\prime=J$. 
A decrease of the inter-layer interaction strength, $J^\prime<J$, decouples the two spins in the pair, meaning that they are more likely to take different states.
Consequently, we might expect the enhancement of infinite avalanches
with reducing $J^\prime$
and an increase in the size of the multi-valued regime for the solution of $\langle m\rangle$. 

\begin{figure}[!tbp]
\includegraphics[width=8.0cm]{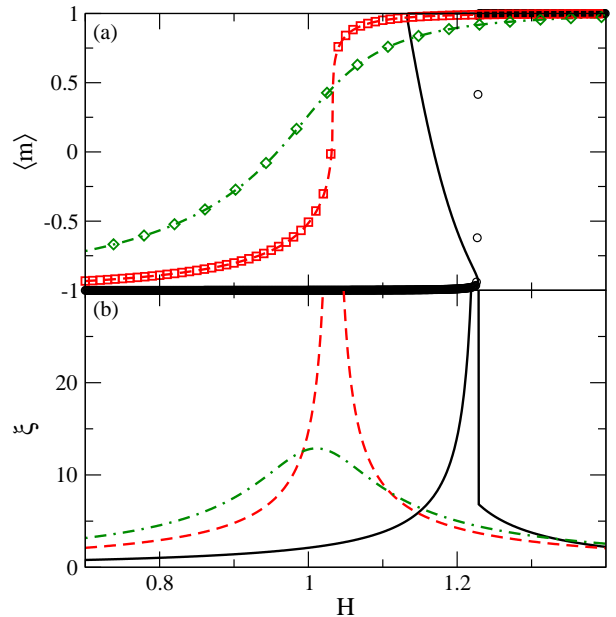}
\caption{Panel (a): Lower half of the hysteresis loops for a $q=3$ bi-layered Bethe lattice with an inter-layer link strength of $J^\prime=0.5122J$. 
The value of $\langle m\rangle$ is plotted against $H$ for disorder above criticality (dot-dashed, $\Delta=1.5$), at criticality (dashed, $\Delta=1.212$) and below criticality at the location of the maximum size of the multivalued region (solid, $\Delta = 0.70$). 
Symbols represent numerical results averaged across $10^2$ realisations of a system of size $N=2\times 10^7$.
In panel (b), the correlation length is plotted {\it vs} external field for the same system and values of the parameters as in panel (a).
\label{fig:InterLayerHysteresis}}
\end{figure}

\begin{figure}[!tbp]
\includegraphics[width=9.0cm]{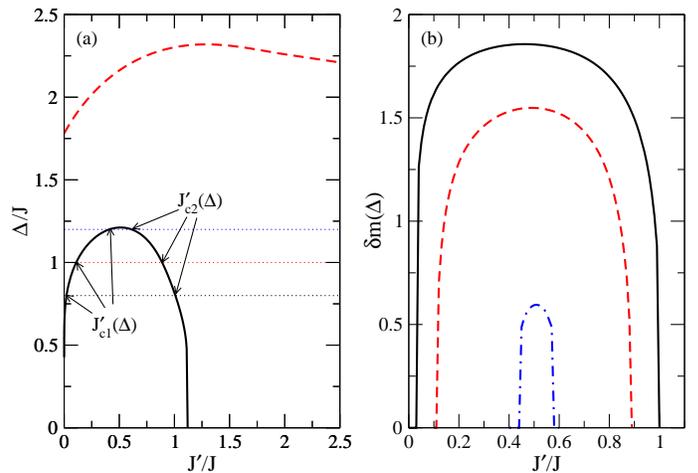}
\caption{Panel (a): Critical disorder $\Delta_{\text{c}}/J$ as a function of the relative strength of the inter-layer links, $J^\prime/J$. 
Solid and dashed lines refer to a bi-layered Bethe lattice with $q=3$ and $q=4$, respectively.
Panel (b): Size of jump in magnetisation, $\delta m(\Delta)$, as a function of $J^\prime/J$ for a bi-layered Bethe lattice with $q=3$ for several degrees of disorder, $\Delta/J=0.8$ (solid line), $\Delta/J=1.0$ (dashed line) and $\Delta/J=1.2$ (dot-dashed line).
The dotted lines in panel (a) correspond to the values of $\Delta/J$ for the curves plotted in panel (b).
\label{fig:CritWithJ}}
\end{figure}

We have solved the self-consistent Eqs.~\eqref{eq:AMatrix}-\eqref{eq:recursiveBilayer} for magnetisation as a function of external field for $J^\prime\neq J$ (see Fig.~\ref{fig:InterLayerHysteresis}(a)) 
and established the phase diagram for this case (see dashed line in Fig.~\ref{fig:phaseDiagram}(a),(b)).
Additionally, we have calculated the correlation length $\xi$ as a function of $H$ (see Fig.~\ref{fig:InterLayerHysteresis}(b)) and confirmed that the critical exponents are not altered from their values when $J^\prime=J$.
Repeating such analysis for several values of $J^\prime$, we have found that the dependence of the critical degree of disorder $\Delta_{\text{c}}$ on $J^\prime$ 
is non-monotonic and reaches a maximum value $\Delta_{\text{c}}^{\text{max}}/J=1.212$ at $J^\prime/J=0.5122$ (see solid line in Fig.\ref{fig:CritWithJ}(a)).
At this point the size of the multivalued region is large in comparison to that for $J^\prime/J=1$ (c.f. regions bounded by a solid and dashed curves in Figs.~\ref{fig:phaseDiagram}(a),(b)).
At low inter-layer link strength, avalanches pass down each layer independently, and the system approaches the behaviour of a $q=3$ single-layered Bethe lattice, i.e. the discontinuity does not exist at $J^\prime=0$. 
The numerical analysis seems to suggest that there will be a jump for any non-zero inter-layer link strength, $J^\prime>0$, although the narrow width of the multivalued region for both $J^\prime\to 0$ and $J^\prime \agt 1.1J$ makes numerical studies imprecise for these values of $J^\prime$. 
In order to avoid these numerical instabilities, we have analysed the dependence of 
$\Delta_{\text{c}}$ on $J^\prime$ for a $q=4$ bi-layered Bethe lattice 
and have found a similar peaked shaped curve (see dashed line in Fig.~\ref{fig:CritWithJ}(a)). 
For this system, the critical disorder has a non-zero limiting value for both small and large $J^\prime$, which can be related to the fact that the $q=4$ single-layer Bethe lattice exhibits a disorder induced phase transition at a non-zero value of disorder.

At a fixed value of $\Delta<\Delta_{\text{c}}^{\text{max}}$, the dependence of the size of the discontinuity in magnetisation, $\delta m(\Delta)$, on $J^\prime$ is not monotonic. Fig.~\ref{fig:CritWithJ}(b) shows that $\delta m(\Delta)$ takes a finite value in the interval $J^\prime_{\text{c}1}(\Delta)<J^\prime<J^\prime_{\text{c}2}(\Delta)$ and is zero outside this region.  
The fact that $\delta m(\Delta)$ can decrease for increasing $J^\prime$ might seem to be counter-intuitive.
However, such a dependence can be understood in terms of the mechanisms of the infinite avalanche presented above.
Indeed, decreasing the inter-layer interaction strength allows the two spins in a pair to take different values, reducing the probability that the outgoing path for the avalanche is blocked.

\section{Conclusions}\label{sec:Conclusions}

In conclusion, we have demonstrated a method to calculate the magnetisation hysteresis loop 
for the random-field Ising model at zero-temperature defined on topologies more complex than a Bethe lattice (i.e. containing topological loops). 
In particular, we have analysed a bi-layered Bethe lattice~\cite{Lyra1992,Hu1997}.
Our main findings are:
(i) an exact implicit solution for magnetisation of a $q=3$ bi-layered Bethe lattice, expressed by a set of $5$ simultaneous self-consistent equations, which can be solved numerically;
(ii) the mechanism by which blocking effects that prevent the presence of infinite avalanches in $q=3$ Bethe lattice are avoided by the introduction of a second layer;
(iii) the effects of the inter-layer interaction strength are studied and shown to be able to increase the range of degrees of disorder in which magnetisation exhibits a discontinuity as a function of external field;
(iv) a reduction in the inter-layer interaction strength can increase the size of the jump in magnetisation, i.e. enhance the size of the infinite avalanche;
(v) the correlation length as a function of external field exhibits a power-law divergence at the critical point with the same critical exponents as those for a $q>3$ single-layered Bethe lattice.

The bi-layered Bethe lattice has local loops.
We have demonstrated how to take into account the effect of these loops exactly in the calculation of the mean magnetisation and correlation functions.
In addition we have described the mechanism which allows the infinite avalanche to grow in the presence of these local loops.
As such this analysis is a step forwards in understanding the mechanism by which an infinite avalanche occurs in more complex and realistic lattices.

The method for evaluation of magnetisation introduced in this paper is based on recursion relations and can be relatively straightforwardly generalised to multi-layer Bethe lattices and to similar topologies but with more than two spin states per node.
The mechanism of infinite avalanches described in this paper applicable to Bethe lattices which can only be embedded in hyperbolic topology~\cite{Soderberg1993} could serve as a basis for an understanding of similar processes on systems embedded in a Euclidean space.

\section{Acknowledgements}

TPH thanks the UK EPSRC for financial support.

\appendix
\section{Categories}\label{sec:Categories}

\begin{figure*}[!tpb]
\includegraphics[width=15cm]{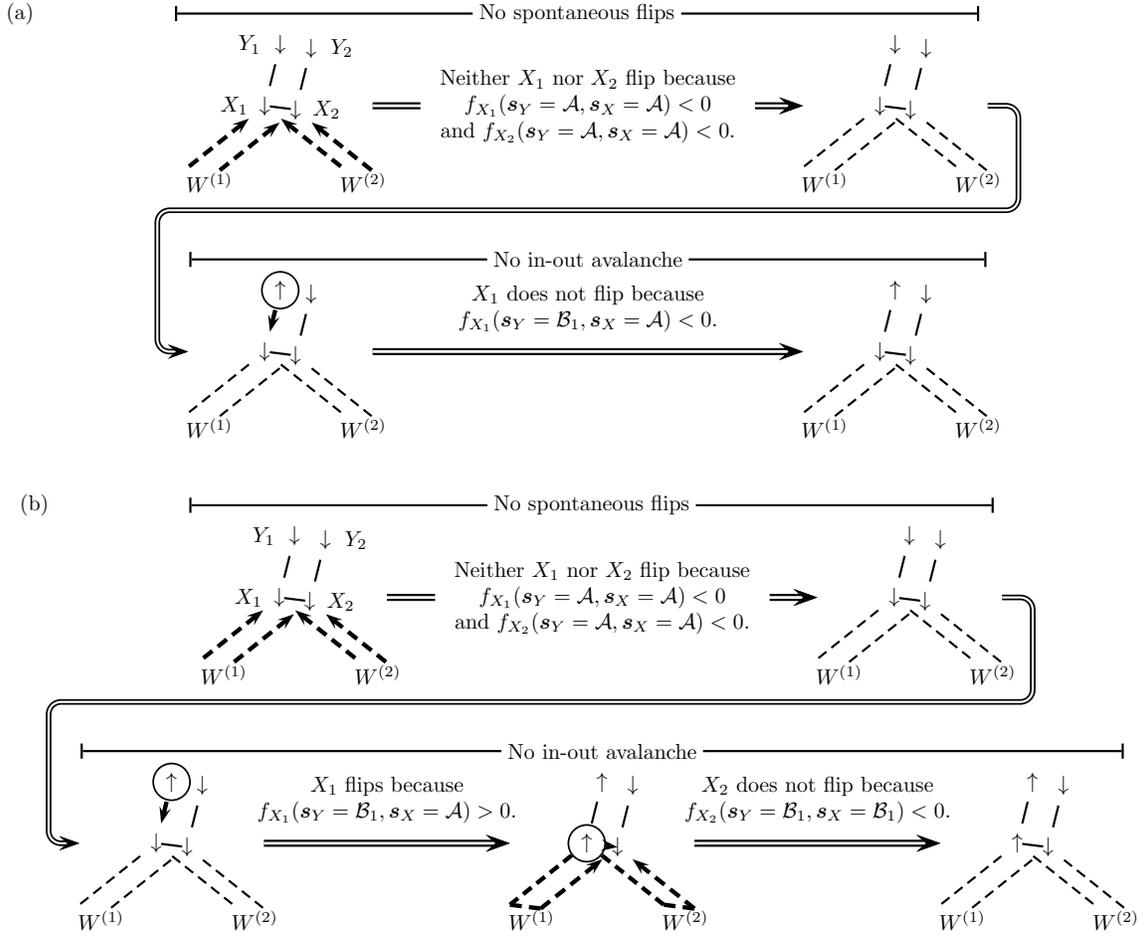}
\caption{Panels (a) and (b) display the two possible sequence of spin
  flips leading to the state function ${\bm s}_X({\bm s}_Y)$ being in category
  $C_1$ for which avalanches emerging from the branch below spin
    pair $X$ do not lead to a spontaneous avalanche at $X$. 
Double arrows represent the order in which events take place.
Bold solid arrows represent the sequence of flips in an avalanche.
Bold dashed arrows represent that there can be an avalanche from the spin pair below.
Bold dashed arrows which go down and then up represent that in-out avalanches 
can occur in the generation below.
Circle around up arrow represents an intermediate state in which a
spin flips.
The light dashed lines represent the links to the spin pairs ${\bm s}_{W^{(b)}}$ which are not shown in the diagram.
\label{fig:categories1}}
\end{figure*}
\begin{figure*}[!tbp]
\includegraphics[width=14cm]{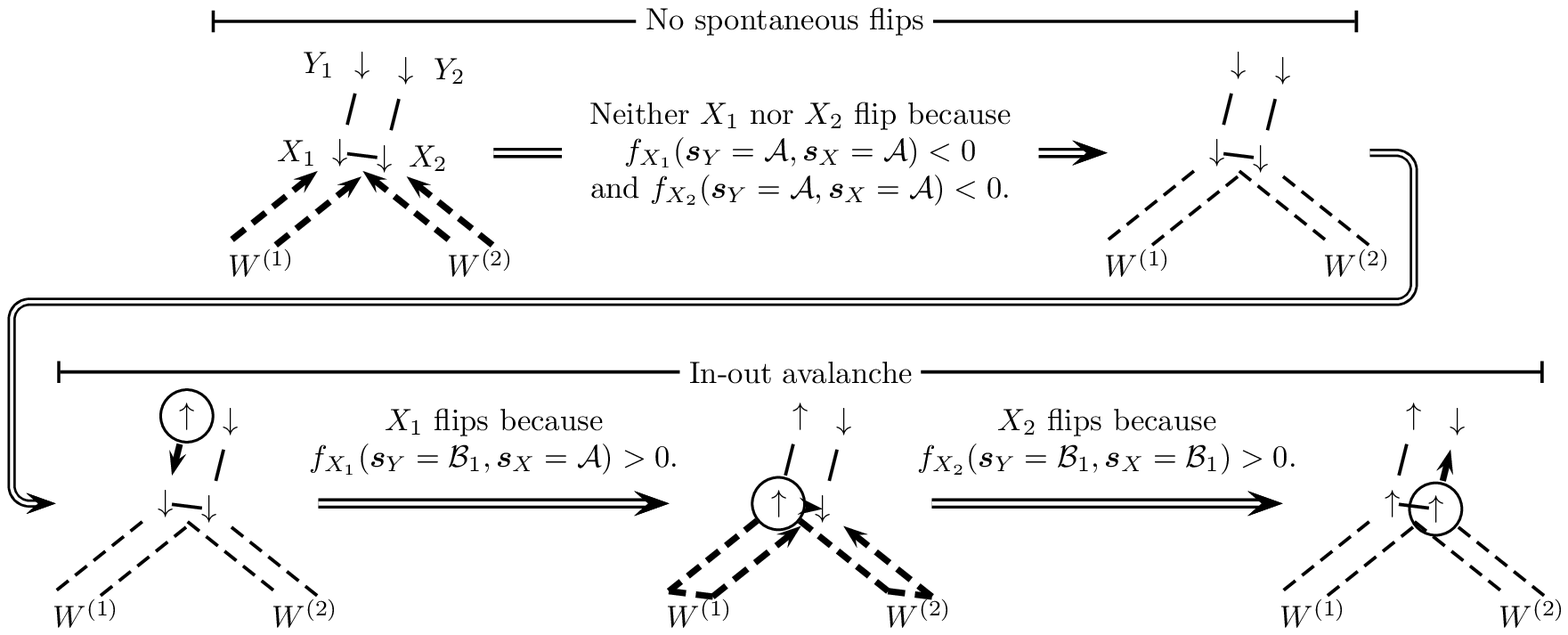}
\caption{Sequence leading to a state function ${\bm s}_X({\bm s}_Y)$ of category $C_2$. Same notation as previous figure.
\label{fig:categories2}}
\end{figure*}
\begin{figure*}[!tbp]
\includegraphics[width=14cm]{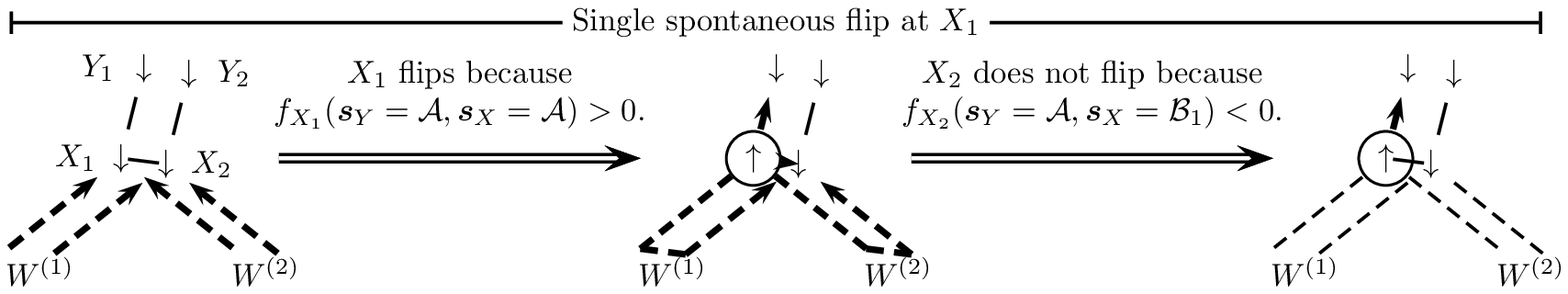}
\caption{Sequence leading to a state function ${\bm s}_X({\bm s}_Y)$ of category $C_3$. Same notation as previous figure.
\label{fig:categories3}}
\end{figure*}
\begin{figure*}[!tbp]
\includegraphics[width=14cm]{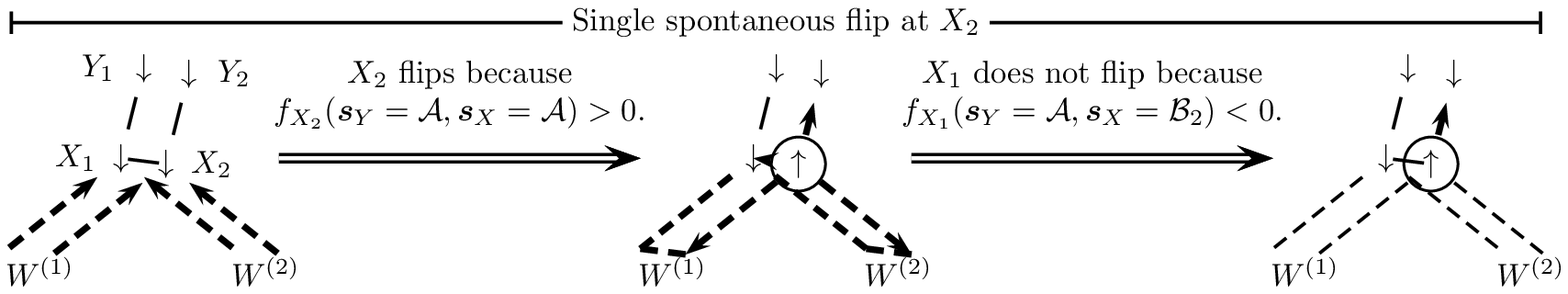}
\caption{Sequence leading to a state function ${\bm s}_X({\bm s}_Y)$ of category $C_4$. Same notation as previous figure.
\label{fig:categories4}}
\end{figure*}
\begin{figure*}[!tbp]
\includegraphics[width=15cm]{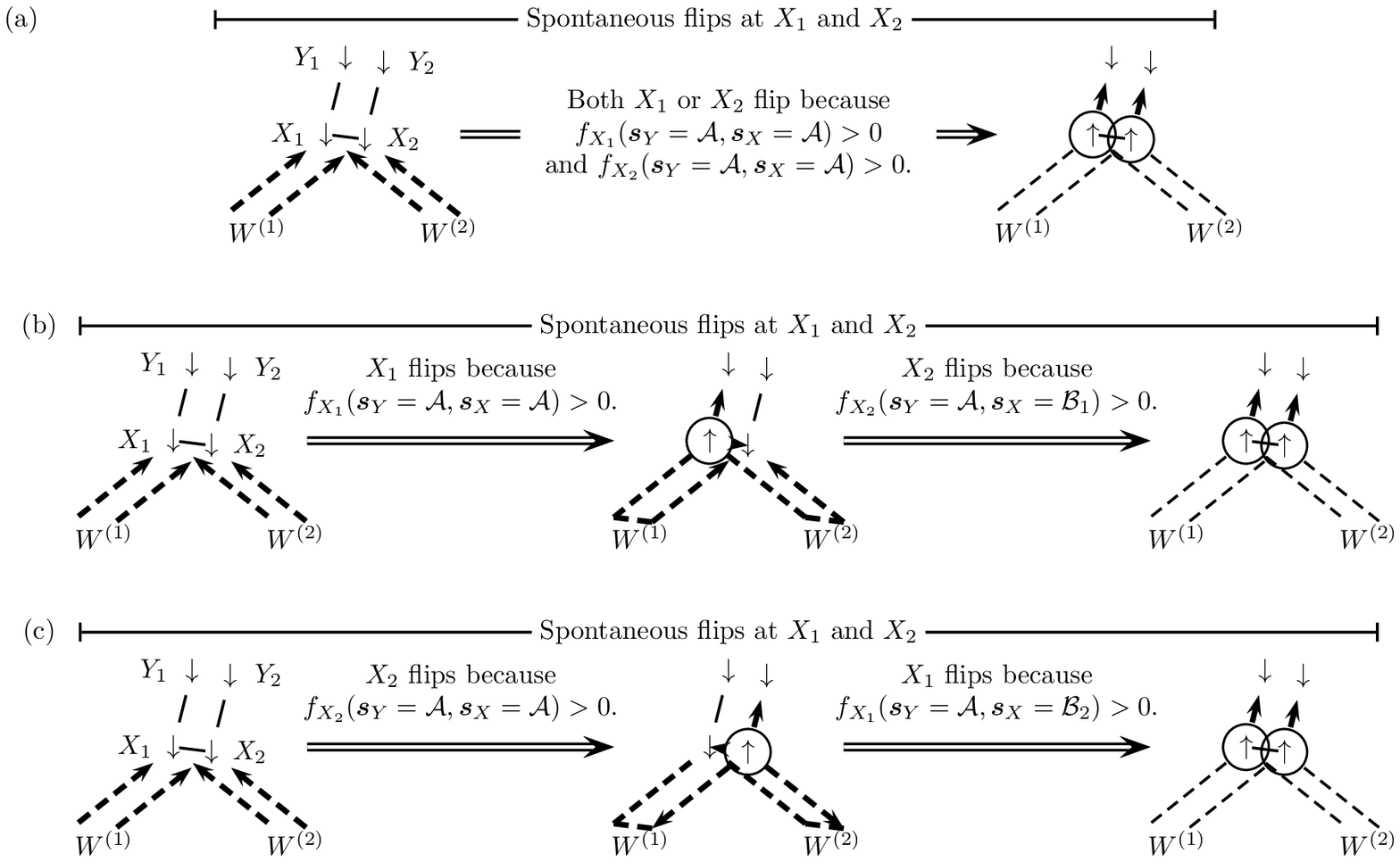}
\caption{Panels (a),(b) and (c) show three possible sequences leading to a state function ${\bm s}_X({\bm s}_Y)$ of category $C_5$. Same notation as previous figure.
In Panels (b) and (c), the upward pointing bold arrow is shown for both the second and the last state.
This represents the fact that the spins $Y_1$ and $Y_2$ which this interaction affects cannot change state until the spin pair $X$ is in a stable state, so that the spin pair $Y$ may be in an unstable state but not flip before the final step shown.
\label{fig:categories5}}
\end{figure*}

As mentioned in Sec.~\ref{sec:Bilayer}, the state function ${\bm s}_X({\bm s}_Y)$ describing the state of a spin pair ${\bm s}_X$ in generation $g$ in terms of the state of the spin pair ${\bm s}_Y$ in the generation $g+1$ can be categorised into $5$ forms. 
In this appendix we give a more detailed description of the categories. 
They are formally defined as follows 
(see table \ref{Table_1} and Figs.~\ref{fig:categories1}-\ref{fig:categories5}),
\begin{eqnarray}
C_1&=&\{{\bm s}_{X}({\bm s}_Y)|({\bm s}_{X}({\bm s}_Y={\cal A})={\cal A}\nonumber\\&&~~~~~~~~~\cap {s}_{X_{2}}({\bm s}_Y={\cal B}_1)=-1)\}\nonumber\\ 
C_2&=&\{{\bm s}_{X}({\bm s}_Y)|({\bm s}_{X}({\bm s}_Y={\cal A})={\cal A}\nonumber\\&&~~~~~~~~~\cap {s}_{X_{2}}({\bm s}_Y={\cal B}_1)=+1)\}\nonumber\\
C_3&=&\{{\bm s}_{X}({\bm s}_Y)|{\bm s}_{X}({\bm s}_Y={\cal A})={\cal B}_1\} \nonumber\\
C_4&=&\{{\bm s}_{X}({\bm s}_Y)|{\bm s}_{X}({\bm s}_Y={\cal A})={\cal B}_2\} \nonumber\\
C_5&=&\{{\bm s}_{X}({\bm s}_Y)|{\bm s}_{X}({\bm s}_Y={\cal A})={\cal C}\}~.\label{eq:categories}
\end{eqnarray}

The category $C_1$ contains all functional forms that lead to no
spontaneous avalanches emerging from spin pair ${\bm s}_{X}$ and the branch below it
(the state ${\bm s}_{X}({\bm s}_Y={\cal A})={\cal A}$), i.e. the fields at
both $s_{X_1}$ and $s_{X_2}$ are negative when both spins in pair ${\bm s}_Y$ are in the down-state, ${\bm s}_Y={\cal A}$ (see Fig.~\ref{fig:categories1}(a)-(b)).
Additionally, functional forms falling into category $C_1$ indicate
that any avalanche propagating into the branch along layer $\ell=1$,
will not propagate back out again along 
layer $\ell=2$, i.e. ${\bm s}_{X_2}({\bm s}_Y={\cal B}_1)=-1$, either
because the field at $s_{X_1}$ remains negative when ${\bm s}_Y={\cal B}_1$
(see Fig.~\ref{fig:categories1}(a)) or because the local field
$f_{X_1}$ at spin $s_{X_1}$ becomes
positive but the field $f_{X_2}$ at 
spin $s_{X_2}$ remains negative (see Fig.~\ref{fig:categories1}(b)). 

Category $C_2$ contains functional forms leading to no spontaneous avalanche emerging from the branch (i.e. the fields at $s_{X_1}$ and $s_{X_2}$ are both negative; see Fig.~\ref{fig:categories2}). 
Category $C_2$ also indicates that an in-out avalanche can propagate into the branch along layer $\ell=1$ (from spin $s_{Y_1}$ to spin $s_{X_1}$) and pass to spin $s_{X_2}$ in the layer $2$ (increasing the local field at spin $s_{Y_{2}}$). 
This occurs when the field at spin $s_{X_1}$ becomes positive and its flip causes the field at spin $s_{X_{2}}$ to become positive.

Functional forms in category $C_3$ or $C_4$ lead to a spontaneous flip of spins in either layer $1$ or $2$,
respectively, but not simultaneously in both layers (see Figs.~\ref{fig:categories3}-\ref{fig:categories4}). 
For $C_3$, this occurs when the field at spin $s_{X_1}$ is positive and the field at $s_{X_2}$ is negative even after $X_1$ has flipped, or vice versa in the case of $C_4$.

Category $C_5$ indicates that both ${s}_{X_1}$ and $s_{X_2}$ flip spontaneously, either because the field at both spins is positive when both ${s}_{X_1}$ and $s_{X_2}$ are down (Fig.~\ref{fig:categories5}(a)), or because the field at spin $s_{X_1}$ is positive and its flip $s_{X_1}$ causes the field at spin $s_{X_2}$ to become positive (Fig.~\ref{fig:categories5}(b)) or vice versa (Fig.~\ref{fig:categories5}(c)).

\section{Derivation of self-consistent equation for the probabilities, $P_i^{(g)}$ [Eq.~\eqref{eq:tensorRecursion}]}\label{sec:SelfConsistent}

The recursion relation between the probabilities ${P}^{(g)}_i$ of ${\bm s}_X({\bm s}_Y)$ being in category $C_i$ and the probabilities ${P}^{(g-1)}_j$ of the functions $\{{\bm s}_{W^{(b)}}({\bm s}_X)\}$ being in categories $C_j$ can be found in the following way.
For convenience, definitions of the possible categories of the function ${\bm s}_X({\bm s}_Y)$ in terms of the local fields at spins ${s}_{X_1}$ and $s_{X_2}$, matching the descriptions given in App.~\ref{sec:Categories}, can be formally written as,
\begin{widetext}
\begin{equation}
{\bm s}_X({\bm s}_Y) \in 
\begin{cases}
C_1~&\text{if}~~[f_{X_1}({\bm s}_Y={\cal B}_1,{\bm s}_X={\cal A})<0 \cap f_{X_2}({\bm s}_Y={\cal B}_1,{\bm s}_X={\cal A})<0]~~~~\\
&~~~~~~~\cup [f_{X_1}({\bm s}_Y={\cal A},{\bm s}_X={\cal A})<0<f_{X_1}({\bm s}_Y={\cal B}_1,{\bm s}_X={\cal A}) \cap f_{X_2}({\bm s}_Y={\cal B}_1,{\bm s}_X={\cal B}_1)<0]~,  \\
C_2~&\text{if}~~f_{X_1}({\bm s}_Y={\cal A},{\bm s}_X={\cal A})<0<f_{X_1}({\bm s}_Y={\cal B}_1,{\bm s}_X={\cal A}) ~~~~\\
&~~~~~~~\cap f_{X_2}({\bm s}_Y={\cal A},{\bm s}_X={\cal A})<0<f_{X_2}({\bm s}_Y={\cal B}_1,{\bm s}_X={\cal B}_1)~,\\
C_3~&\text{if}~~f_{X_1}({\bm s}_Y={\cal A},{\bm s}_X={\cal A})>0 \cap f_{X_2}({\bm s}_Y={\cal A},{\bm s}_X={\cal B}_1)<0~,\\
C_4~&\text{if}~~f_{X_2}({\bm s}_Y={\cal A},{\bm s}_X={\cal A})>0 \cap f_{X_1}({\bm s}_Y={\cal A},{\bm s}_X={\cal B}_2)<0~,\\
C_5~&\text{if}~~[f_{X_1}({\bm s}_Y={\cal A},{\bm s}_X={\cal A})>0 \cap f_{X_2}({\bm s}_Y={\cal A},{\bm s}_X={\cal B}_1)>0]~~~~\\
&~~~~~~~\cup[f_{X_2}({\bm s}_Y={\cal A},{\bm s}_X={\cal A})>0 \cap f_{X_1}({\bm s}_Y={\cal A},{\bm s}_X={\cal B}_2)>0].
\end{cases}\label{eq:sxsyCat}
\end{equation}
The above definition does not explicitly contain the dependence on the state functions, $\{{\bm s}_{W^{(b)}}({\bm s}_X)\}$, of the spin pairs ${\bm s}_{W^{(b)}}$ ($1\le b\le q-1$) in the generation below. 
These state functions are adsorbed into the expression for local fields $f_{X_1}({\bm s}_Y={\cal A},{\bm s}_X={\cal A})$, which in fact do depend on $\{{\bm s}_{W^{(b)}}({\bm s}_X)\}$, according to the following equation,
\begin{equation}
f_{X_\ell}({\bm s}_Y,{\bm s}_X;\{{\bm s}_{W^{(b)}}({\bm s}_X)\},h_{X_\ell})=J{\bm s}_{Y_\ell}+J\sum_{b=1}^{q-1}{s}_{W^{(b)}_\ell}({\bm s}_X)+H+h_{X_\ell}+J^\prime {\bm s}_{X_{3-\ell}}~.\label{eq:locFieldX}
\end{equation}
In Eq.~\eqref{eq:locFieldX}, $X_{3-\ell}$ refers to the spin in layer $3-\ell$ which is linked to spin $X_\ell$ in layer $\ell=1$ or $2$. 
The random function $f_{X_\ell}({\bm s}_Y,{\bm s}_X;\{{\bm s}_{W^{(b)}}({\bm s}_X)\},h_{X_\ell})$ accounts for the effects of the random fields at spin pair $X$ and all the avalanches which randomly emerge from lower generations.
Eq.~\eqref{eq:locFieldX} can be rewritten in terms the categories of either $\{{\bm s}_{W^{(b)}}({\bm s}_X)\}$ or $\{{\bm s}_{W^{(b)}}^\prime({\bm s}_X^\prime)\}$ as follows,
\begin{eqnarray}
f_{X_1}({\bm s}_Y,{\bm s}_X={\cal A};\{{\bm s}_{W^{(b)}}({\bm s}_X)\},h_{X_1})&=&J{s}_{Y_1}+H+h_{X_1}+J(2n_3+2n_5-(q-1))-J^\prime\nonumber\\
&=&J{s}_{Y_1}+H+h_{X_1}+J(2n_4^\prime+2n_5^\prime-(q-1))-J^\prime\nonumber\\
f_{X_2}({\bm s}_Y,{\bm s}_X={\cal A};\{{\bm s}_{W^{(b)}}({\bm s}_X)\},h_{X_2})&=&J{s}_{Y_2}+H+h_{X_2}+J(2n_4+2n_5-(q-1))-J^\prime\nonumber\\
&=&J{s}_{Y_2}+H+h_{X_2}+J(2n_3^\prime+2n_5^\prime-(q-1))-J^\prime\nonumber\\
f_{X_1}({\bm s}_Y,{\bm s}_X={\cal B}_2;\{{\bm s}_{W^{(b)}}({\bm s}_X)\},h_{X_1})&=&J{s}_{Y_1}+H+h_{X_1}+J(2n_2^\prime+2n_4^\prime+2n_5^\prime-(q-1))+J^\prime\nonumber\\
f_{X_2}({\bm s}_Y,{\bm s}_X={\cal B}_1;\{{\bm s}_{W^{(b)}}({\bm s}_X)\},h_{X_2})&=&J{s}_{Y_2}+H+h_{X_2}+J(2n_2+2n_4+2n_5-(q-1))+J^\prime~,\label{eq:specFields}
\end{eqnarray}
where we have defined $n_i$ and $n^\prime_i$ as the number of the functions ${\bm s}_{W^{(b)}}({\bm s}_X)$ and ${\bm s}_{W^{(b)}}^\prime({\bm s}_X^\prime)$, respectively, in category $C_i$.
As explained in the Sec.~\ref{sec:Bilayer}, the functions ${\bm s}_{W^{(b)}}^\prime({\bm s}_X^\prime)$ are the state functions that spin pair $W^{(b)}$ would have if all of the spins in layer $\ell=1$ were swapped with their copies in layer $\ell=2$.
Substitution of Eq.~\eqref{eq:specFields} into Eq.~\eqref{eq:sxsyCat} gives a set of conditional expressions for the categories of ${\bm s}_X({\bm s}_Y;h_{X_1},h_{X_2},\{{\bm s}_{W^{(b)}}({\bm s}_X)\})$ in terms of both the random fields $h_{X_1}$, $h_{X_2}$ and categories of ${\bm s}_{W^{(b)}}({\bm s}_X)$ and ${\bm s}_{W^{(b)}}^\prime({\bm s}_X^\prime)$. 
The probability distribution of the category of ${\bm s}_X({\bm s}_Y)$ can then be written as,
\begin{eqnarray}
P(C_i,g)=P({\bm s}_X({\bm s}_Y)\in C_i)&=&\int\limits_{h_{X_1}=-\infty}^{\infty}\int\limits_{h_{X_2}=-\infty}^{\infty}\sum_{n_1=0}^{q-1}\ldots\sum_{n_5=0}^{q-1}\sum_{n^\prime_1=0}^{q-1}\ldots\sum_{n^\prime_5=0}^{q-1}I_{C_i}\left[{\bm s}_X({\bm s}_Y;h_{X_1},h_{X_2},\{{\bm s}_{W^{(b)}}({\bm s}_X)\})\right]\nonumber\\&&\times P\left(\{n_i\},\{n^\prime_i\}\right)\rho(h_{X_1})\rho(h_{X_2})\text{d}h_{X_1}\text{d}h_{X_2}~,\label{eq:recurIntegrals}
\end{eqnarray}
where $I_{C_i}[S(S^\prime)]$ is the indicator function, defined by $I_{C_i}[S(S^\prime)]=1$ for $S(S^\prime)\in C_i$ and $0$ otherwise.
In Eq.~\eqref{eq:recurIntegrals}, the value of $P\left(\{n_i\},\{n^\prime_i\}\right)$ represents the joint probability distribution of the numbers $\{n_i;\, i=1,2,\dots,5\}$ and $\{n^\prime_i;\, i=1,2,\dots,5\}$ that take values given by the function $n_{i}(j,\ldots,l)$  (cf. Eq.~\eqref{eq:numInCat}). 
Performing the integral over the random fields in Eq.~\eqref{eq:recurIntegrals} results in the following,
\begin{eqnarray}
P(C_1,g)&=&\sum_{n_1=0}^{q-1}\ldots\sum_{n_5=0}^{q-1}P\left(\{n_i\}\right) \left[(1-p_{(n_3+n_5),0})(1-p_{(n_4+n_5+1),0})+(1-p_{(n_2+n_3+n_5),1})(p_{(n_4+n_5+1),0}-p_{(n_4+n_5),0}) \right]\nonumber\\
P(C_2,g)&=&\sum_{n_1=0}^{q-1}\ldots\sum_{n_5=0}^{q-1}P\left(\{n_i\}\right)(p_{(n_2+n_3+n_5),1}-p_{(n_3+n_5),0})(p_{(n_4+n_5+1),0}-p_{(n_4+n_5),0})\nonumber\\
P(C_3,g)&=&\sum_{n_1=0}^{q-1}\ldots\sum_{n_5=0}^{q-1}P\left(\{n_i\}\right)(1-p_{(n_2+n_4+n_5),1})p_{(n_3+n_5),0}\nonumber\\
P(C_4,g)&=&\sum_{n^\prime_1=0}^{q-1}\ldots\sum_{n_5^\prime=0}^{q-1}P\left(\{n^\prime_i\}\right)p_{(n^\prime_3+n^\prime_5),0}(1-p_{(n_2^\prime+n^\prime_4+n^\prime_5),1})\nonumber\\
P(C_5,g)&=&\sum_{n_1=0}^{q-1}\ldots\sum_{n_5=0}^{q-1}P\left(\{n_i\}\right)p_{(n_3+n_5),0}p_{(n_2+n_4+n_5),1}+\sum_{n^\prime_1=0}^{q-1}\ldots\sum_{n_5^\prime=0}^{q-1}P\left(\{n^\prime_i\}\right)(p_{(n_2^\prime+n_4^\prime+n_5^\prime),1}-p_{(n_4^\prime+n_5^\prime),0})p_{(n_3^\prime+n_5^\prime),0}~,\label{eq:AMatDeriv}\nonumber\\
\end{eqnarray}
\end{widetext}
where $P\left(\{n_i\}\right)$ and $P\left(\{n^\prime_i\}\right)$ are the distributions of $\{n_i\}$ and $\{n^\prime_i\}$, respectively.
Owing to the symmetry between the layers, the distributions $P(\{n_i\})$ and $P(\{n^\prime_i\})$ should be the same and the primes can be dropped wherever they appear in Eq.~\eqref{eq:AMatDeriv}.
Since the categories, $C_i$, of each of the functions ${\bm s}_{W^{(b)}}({\bm s}_X)$ are distributed independently according to $P(C_i,g)$, the function $P\left(\{n_i\}\right)$ can be written as,
\begin{eqnarray}
P\left(\{n_i\}\right)&=&\sum_{j=1}^5\ldots\sum_{l=1}^5P(C_j,g-1)\ldots P(C_l,g-1)\nonumber\\&&\times \prod_{i=1}^5\delta_{n_{i},n_{i}(j,\ldots,l)}~.\nonumber\\\label{eq:ProbTensor}
\end{eqnarray}
Eq.~\eqref{eq:ProbTensor} can then be substituted into Eq.~\eqref{eq:AMatDeriv}, revealing the tensor equation~\eqref{eq:tensorRecursion} in the main text.

\section{Probability of Root Spin Flipping in bi-layered Bethe lattice}\label{sec:rootSite}

The root spin $R_1$ will flip on the last iteration step if either the local field at that spin is positive at the start of that step, 
or if (a) the field at the other root spin is positive, 
causing it to flip and (b) the field at $R_1$ is positive once $s_{R_2}$ has flipped, i.e. 
\begin{equation}
s_{R_1}=\begin{cases}
+1 & \text{if}~f_{R_1}({\bm s}_R={\cal A};\{{\bm s}_{W^{(b)}}({\bm s}_R)\},h_{R_1})>0 \\
& \cup[f_{R_2}({\bm s}_R={\cal A};\{{\bm s}_{W^{(b)}}({\bm s}_R)\},h_{R_2})>0 \\
& ~\cap f_{R_1}({\bm s}_R={\cal B}_2;\{{\bm s}_{W^{(b)}}({\bm s}_R)\},h_{R_1})>0]\\
-1~, & \text{otherwise}~.
\end{cases}\label{eq:RootFlipCond}
\end{equation}
Here, the local fields at the root spins can be written as,
\begin{eqnarray}
&&f_{R_\ell}({\bm s}_R;\{{\bm s}_{W^{(b)}}({\bm s}_R)\},h_{R_\ell})\nonumber\\&&~~~~=J\sum_{b=1}^{q}{\bm s}_{W^{(b)}}({\bm s}_R)+H+h_{R_\ell}+J^\prime s_{R_{3-\ell}}~,\label{eq:locFieldCentral}
\end{eqnarray}
where $\{{\bm s}_{W^{(b)}}({\bm s}_R)\}$ ($1\le b\le q$) describes the states of the spins in generation ${\bar g}-1$ in terms of the state of the root spins.
Eq.~\eqref{eq:locFieldCentral} can be rewritten in terms of the numbers $m_i$ ($1\le i\le 5$) of the functions ${\bm s}_{W^{(b)}}^\prime({\bm s}^\prime_R)$ (the layer-swapped versions of ${\bm s}_{W^{(b)}}({\bm s}_R)$; see Sec.~\ref{sec:Bilayer}) in each category $C_i$,
\begin{eqnarray}
&&f_{R_1}({\bm s}_R={\cal A};\{{\bm s}_{W^{(b)}}({\bm s}_X)\},h_{X_1})\nonumber\\
&&=H+h_{R_1}+J(2m_4+2m_5-(q-1))-J^\prime\nonumber\\
&&f_{R_2}({\bm s}_R={\cal A};\{{\bm s}_{W^{(b)}}({\bm s}_X)\},h_{X_1})\nonumber\\
&&=H+h_{R_2}+J(2m_3+2m_5-(q-1))-J^\prime\nonumber\\
&&f_{R_1}({\bm s}_R={\cal B}_2;\{{\bm s}_{W^{(b)}}({\bm s}_X)\},h_{X_1})\nonumber\\
&&=H+h_{R_1}+J(2m_2+2m_4+2m_5-(q-1))+J^\prime~.\nonumber\\\label{eq:RootLocalFieldParticular}
\end{eqnarray}
The value of ${s}_{R_1}$ is a random number depending on $h_{R_1}$, $h_{R_2}$ and $\{m_i\}$.
The probability that ${s}_{R_1}=+1$ can be written,
\begin{eqnarray}
P({s}_{R_1}=+1)&=&\int\limits_{h_{R_1}=-\infty}^{\infty}\int\limits_{h_{R_2}=-\infty}^{\infty}\sum_{m_1=0}^q\ldots\sum_{m_5=0}^q \delta_{s_{R_1},+1}\nonumber\\&\times&
P(\{m_i\})\rho(h_{R_1})\rho(h_{R_2})~\text{d}h_{R_1}\text{d}h_{R_2}~.\nonumber\\\label{eq:rootBilayer}
\end{eqnarray}
As in the case for the recursive relation given by Eq.~\eqref{eq:recurIntegrals}, substitution of Eqs.~\eqref{eq:RootFlipCond} and~\eqref{eq:RootLocalFieldParticular} into Eq.~\eqref{eq:rootBilayer} leads to a relationship,
\begin{eqnarray}
&&P({s}_{R_1}=+1)=\sum_{m_1=0}^q\ldots\sum_{m_5=0}^qP(\{m_i\})\nonumber\\
&&\times p_{(m_4+m_5),0}+(p_{(m_2+m_4+m_5),1}-p_{(m_4+m_5),0})p_{(m_3+m_5),0}~,\nonumber\\\label{eq:RootSums}
\end{eqnarray}
where $P(\{m_i\})$ can be written as, 
\begin{eqnarray}
P(\{m_i\})&=&\sum_{i=1}^5\sum_{j=1}^5\ldots\sum_{l=1}^5\prod_{t=1}^5\delta_{m_t,m_t(i,\ldots,l)}\nonumber\\&&\times P(C_i,{\bar g}-1)P(C_j,{\bar g}-1)\ldots P(C_l,{\bar g}-1)~.\nonumber\\\label{eq:CentProbTensor}
\end{eqnarray}
In Eq.~\eqref{eq:CentProbTensor} there are now $q$ sums, corresponding to each branch, 
and $m_t(i,\ldots,l)$ gives the number of the subscripts $i,\dots k$ which equal $t$, i.e. $m_t(i,\ldots,l)=\delta_{it}+\delta_{jt}+\dots+\delta_{kt}$.
Substituting the expression given by Eq.~\eqref{eq:CentProbTensor} into Eq.~\eqref{eq:RootSums} gives the tensor relationship Eq.~\eqref{eq:rootBilayerM} in the main text.

\end{document}